\documentstyle[twoside,fleqn,espcrc2,psfig]{article}

\newcommand{\AmS}{{\protect\the\textfont2
  A\kern-.1667em\lower.5ex\hbox{M}\kern-.125emS}}

\title{Standard and exotic interpretations of the atmospheric
neutrino data}

\author{N. Fornengo\address{Dipartimento di Fisica Teorica,
 Universit\`a di Torino and INFN, Sezione di Torino, via
 P. Giuria 1, 10125 Torino, Italy \\
 Instituto de F\'{\i}sica 
 Corpuscular -- C.S.I.C., Departamento de F\'{\i}sica Te\`orica, 
 Universitat de Val\`encia, E-46100 Burjassot, Val\`encia, Spain \\
 {\sl fornengo@to.infn.it, fornengo@flamenco.ific.uv.es}}
 \thanks{Report on the work done in collaboration with M.C. Gonzalez-Garcia
 and J.W.F. Valle.}}

\begin{document}

\begin{abstract}
The present status of some theoretical interpretations of
the atmospheric neutrino deficit is briefly discussed. Specifically,
we show the results for the FC mechanism and for the standard
oscillation hypothesis, both in the active and in the sterile
channels. All these mechanisms are able to fit the present data
to a good statistical level. Among them, the $\nu_\mu \to \nu_\tau$ 
oscillation is certainly the best explanation to the 
atmospheric neutrino deficit, providing a remarkably good 
agreement with the data.
\end{abstract}

\maketitle

When cosmic rays collide with nuclei in the upper atmosphere, they
produce neutrino fluxes, which have been detected 
by several detectors over many years
\cite{sk99,Super-Kamiokande,MACRO,Baksan,Soudan,Kamiokande,Frejus,IMB,Nusex}. 
Even though the
absolute fluxes of atmospheric neutrinos are largely uncertain,
the expected ratio $R(\mu/e)$ of the muon neutrino flux  over the 
electron neutrino flux 
is robust, since it largely cancels out the uncertainties associated
with the absolute fluxes. This ratio has been calculated \cite{fluxes} 
with an uncertainty of less than 5\% over energies
varying from 0.1~GeV to 100~GeV. Since the calculated ratio does not
match the observations, we believe to be facing an anomaly which can
be ascribed to non--standard neutrino properties.

Super-Kamiokande high statistics
observations~\cite{sk99,Super-Kamiokande} indicate that the deficit in the
total ratio $R(\mu/e)$ is due to the number of neutrinos reaching 
the detector at large zenith angles. The $e$-like events do not
present any compelling evidence of a zenith-angle dependent suppression 
while the $\mu$-like event rates are substantially suppressed at large zenith
angles. 

%%%%%%%%%%%%%%%%%%%%%%%%%%%%%%%%%%%%%%%%%%%%%%%%%%%%%%%%%%%%%%%%%%%%%%%%%%%%%%%%%%%%%%%%%%%%%%%
\begin{figure}[t]
\hbox{
\psfig{figure=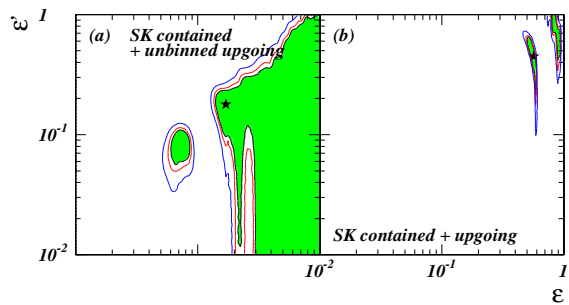,width=4.0in,bbllx=36bp,bblly=260bp,bburx=720bp,bbury=540bp,clip=}
}
\vspace{-30pt}
\caption{FCNC solution to the atmospheric neutrino problem.
Allowed regions for $\epsilon_\nu$ and $\epsilon_\nu^\prime$
for the combination of the 52 kton-yrs Super-Kamiokande data sets:
(a) the binned
contained events are combined with total (unbinned) up-going events;
(b) binned contained and up-going events. The best-fit point for each
case is indicated by a star. The shaded areas refer to 90\%
while the contours stand for 95\% and 99\% C.L.}
\vspace{-30pt}
\end{figure}

\begin{figure*}[t]
\hbox{
\psfig{figure=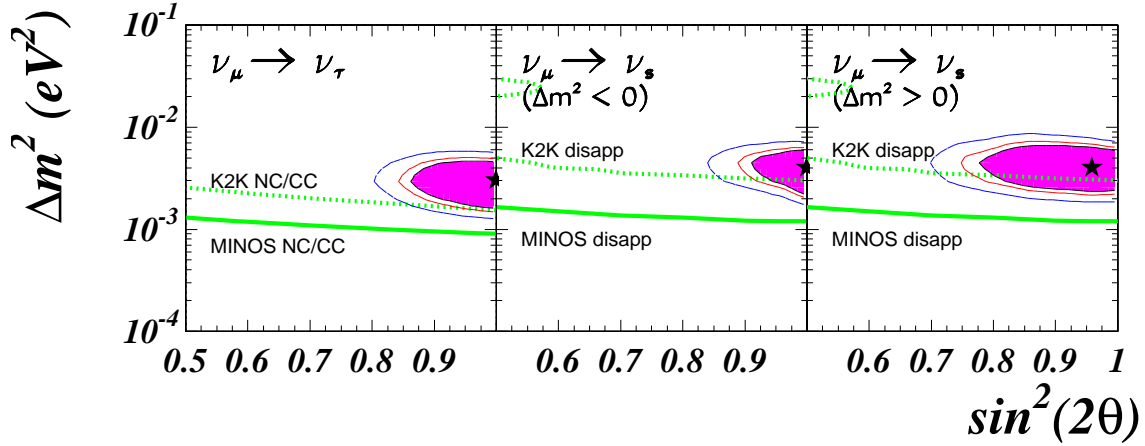,width=8.2in,bbllx=10bp,bblly=315bp,bburx=750bp,bbury=540bp,clip=}
}
\vspace{-30pt}
\caption{Oscillation solution to the atmospheric neutrino problem.
Allowed regions for $\sin^2(2\theta)$
and $\Delta m^2$ obtained from the global fits to all the available
atmospheric neutrino data [1-9]. The three panels refer to the three
oscillation channels: $\nu_\mu \rightarrow \nu_\tau$ (left),
$\nu_\mu \rightarrow \nu_s$ with negative $\Delta m^2$ (center),
and $\nu_\mu \rightarrow \nu_s$ with positive $\Delta m^2$ (right).
The best-fit point for each
case is indicated by a star. The shaded areas refer to 90\% C.L.
while the contours stand for 95\% and 99\% C.L.
On the same plot, expected sensitivities for  K2K and
MINOS are also reported.}
\vspace{-15pt}
\end{figure*}

%%%%%%%%%%%%%%%%%%%%%%%%%%%%%%%%%%%%%%%%%%%%%%%%%%%%%%%%%%%%%%%%%%%%%%%%%%%%%%%%%%%%%%%%%%%%%%%

A simplest explanation for these features comes from the
hypothesis of neutrino masses and neutrino flavour oscillations,
where a $\nu_\mu$ transforms during propagation into a $\nu_\tau$
or, alternatively, a sterile neutrino $\nu_s$. However, alternative (``exotic'')
interpretations to 
the atmospheric neutrino deficit have been proposed. Among others,
flavour changing (FC) neutrino interactions in matter, 
neutrino decay, violation of relativity principles or violation 
of the CPT symmetry (see Ref.\cite{FCNC,OSC} for relevant references).  

In this paper, together with presenting the most updated results
for the ``standard'' solution in terms of neutrino oscillations,
we will also discuss the status of the FC hypothesis. For
the latter case, we will perform the analysis of the latest
52 kton-yrs Super-Kamiokande data \cite{sk99}, while in the
case of the oscillation mechanism we will report on our
global fit to all the available atmospheric neutrino 
data \cite{sk99,Super-Kamiokande,MACRO,Baksan,Soudan,Kamiokande,Frejus,IMB,Nusex}.
More detailed discussions can be found in Ref.\cite{FCNC} for the FC
hypothesis and in Ref.\cite{OSC} for the oscillation mechanism. We
refer to these papers also for a more exhaustive set of references.
Here we briefly recall that we calculate the expected number of 
$\mu$-like and $e$-like contained events 
as $N_\mu= N_{\mu\mu} + N_{e\mu}$ and 
$N_e= N_{ee} +  N_{\mu e}$ where
\begin{eqnarray}
N_{\alpha\beta} &=& n_t T
\int
\frac{d^2\Phi_\alpha}{dE_\nu d(\cos\theta_\nu)} 
\kappa_\alpha(h,\cos\theta_\nu,E_\nu) \nonumber \\
& & P_{\alpha\beta} \frac{d\sigma}{dE_\beta}\varepsilon(E_\beta)
dE_\nu dE_\beta d(\cos\theta_\nu)dh\;
\label{event0}
\end{eqnarray}
where $n_t$ is the number of targets, $T$ is the experiment's running
time, $E_\nu$ is the neutrino energy and $\Phi_\alpha$ is the flux of
atmospheric neutrinos ($\alpha=\mu ,e$); $E_\beta$ is the final
charged lepton energy and $\varepsilon(E_\beta)$ is the detection
efficiency for such charged lepton; $\sigma$ is the neutrino-nucleon
interaction cross section, and $\theta_\nu$ is zenith angle;
$h$ and $\kappa_\alpha$ are geometrical factors \cite{OSC0}.
$P_{\alpha\beta}$ is the conversion probability of $\nu_\alpha \to
\nu_\beta$, which depends on the conversion mechanism. See  
Refs.\cite{FCNC,OSC,OSC0} for the relevant expressions.
For the upgoing muon data we calculate the fluxes as
\begin{equation}
\Phi_\mu(\theta)_{S,T}=\frac{1}{A(L,\theta)}\int 
\frac{d\Phi_\mu(E_\mu,\theta)}{dE_\mu} 
A_{S,T}(E_\mu,\theta)
\end{equation}  
where
\begin{eqnarray}
\frac{d\Phi_\mu}{dE_\mu} 
&=& \int\frac{d\Phi_{\nu_\mu}(E_\nu,\theta)}{dE_\nu} 
P_{\mu\mu} \frac{d\sigma}{dE_{\mu 0}} 
R(E_{\mu 0}, E_\mu) \nonumber \\ 
& & \kappa_\mu(h,\cos\theta_\nu,E_\nu) dE_{\mu 0} dE_\nu dh 
\end{eqnarray}
where $R(E_{\mu 0}, E_\mu)$ is the muon range function,
$A(L,\theta)=A_{S}(E_\mu,\theta)+A_{T}(E_\mu,\theta)$ 
is the projected detector area for internal pathlengths longer than $L$.
$A_{S}$ and $A_{T}$ are the corresponding areas for stopping 
and through-going muon trajectories.

The fitting procedure we adopt is discussed in detail in
\cite{OSC0,OSC}. Here we only recall that we define
a $\chi^2$ function 
\begin{equation}
\chi^2 \equiv \sum_{I,J}
(N_I^{da}-N_I^{th}) \cdot 
(\sigma_{da}^2 + \sigma_{th}^2 )_{IJ}^{-1}\cdot 
(N_J^{da}-N_J^{th}),
\label{chi2}
\end{equation}
where $I$ and $J$ stand for any combination of the experimental data
sets and event-types considered. The error matrices are defined as
$\sigma_{IJ}^2 \equiv \sigma_\alpha(A)\, \rho_{\alpha \beta} (A,B)\,
\sigma_\beta(B)$
where $\rho_{\alpha \beta} (A,B)$ is the correlation matrix. A
detailed discussion of the errors and correlations used in our
analysis can be found in Ref.~\cite{OSC,OSC0}.  
The final step is the minimization of the $\chi^2$ function 
from which we determine the allowed region in the parameter space
as: $\chi^2 \equiv \chi_{min}^2  + 4.6, 6.0, 9.2 $ for 90,95 and 99 \% C.L.

As for the FC mechanism, we report in Table 1 the result of our fits
over the different 52 kton-yrs SK data samples \cite{sk99}. 
The same table also shows our
results for the oscillation interpretation. We notice that the FC
hypothesis is able to fit well all the different data sets, with
statistical confidence comparable to the oscillation cases. 
When a global analysis is performed, the FC hypothesis turns out to be
a worse explanation as compared to oscillation. This is mainly
due to a too strong suppression of the horizonthal thru-going muons \cite{FCNC}.
Nevertheless, the FC mechanism is still acceptable at 90\% C.L. . Fig. 1
shows the allowed regions in the two-parameter space of the FC
mechanism \cite{FCNC}. We can notice that, in order to describe the
data, a somewhat large amount of FC in the neutrino sector is required.

As for the oscillation mechanism, we have peformed a global fit to all
the available atmospheric neutrino 
data: Nusex \cite{Nusex}, IMB \cite{IMB}, Frejus \cite{Frejus}, 
Kamiokande \cite{Kamiokande}, Soudan \cite{Soudan}, Super Kamiokande \cite{sk99}, 
MACRO \cite{MACRO} and Baskan \cite{Baksan}.
Some of our results are shown in Fig. 2 and in Table 2. We see that
all the three oscillation channels describe the data 
to a good statistical level (for details, see \cite{OSC}). 
From the results of the fit, we can conclude that, among the
three possibilities, the $\nu_\mu \to \nu_\tau$ oscillation hypothesis 
turns out to be the current most favourable option.

%%%%%%%%%%%%%%%%%%%%%%%%%%%%%%%%%%%%%%%%%%%%%%%%%%%%%%%%%%%%%%%%%%%%%%%%%%%%%%%%%%%%%%%%%%%%%%%
\begin{table}
\begin{tabular}{|l|r|r|r|r|r|}
\hline
SK data & d.o.f & FC & \multicolumn{3}{c|}{oscillation} \\
%\hline
$\,$ & $\,$  & $\,$ & $(a)$ & $(b)$ & $(c)$  \\ 
\hline
sub-GeV                  &  8  &  2.4  &  2.4  &  2.7  &  2.7  \\
multi-GeV                &  8  &  6.4  &  6.3  &  9.0  &  8.9  \\
contained                & 18  &  9.3  &  8.9  & 12.9  & 12.6  \\
stop-$\mu$               &  3  &  1.0  &  1.3  &  2.4  &  2.3  \\
thru-$\mu$               &  8  & 10.3  & 10.4  & 13.5  & 10.5  \\
\hline
Global                   & 33  & 44.   &  23.5 & 32.6  & 32.2  \\
\hline
\end{tabular}
\vspace{3pt}

{Table 1. Values of $\chi^2_{min}$ for the different SK data sets [1] and
their combinations. For the neutrino oscillations case, 
$(a)$ refers to $\nu_\mu \rightarrow \nu_\tau$, 
$(b)$ to $\nu_\mu \rightarrow \nu_s$ ($\Delta m^2 < 0$)
and $(c)$ to $\nu_\mu \rightarrow \nu_s$ ($\Delta m^2 > 0$).}
\vspace{-30pt}
\end{table}

\begin{table}
\begin{tabular}{|l|r|r|r|r|}
\hline
 $\nu_\mu \rightarrow$  & $\sin^2(2\theta)$ & $\Delta m^2 (\mbox{eV}^2)$ 
       & $\chi^2_{min}$    & d.o.f. \\ 
\hline 
  $\nu_\tau$   & 1.00 & 3.0 $\cdot 10^{-3}$
   & 58.5 & 61 \\
  $\nu_s$ ($-$)  & 1.00 & 4.0 $\cdot 10^{-3}$ 
   & 50.9 & 51 \\
  $\nu_s$ ($+$) & 0.96 & 3.0 $\cdot 10^{-3}$ 
   & 50.4 & 51 \\
\hline
\end{tabular}
\vspace{3pt}

{Table 2. Best fit results for the oscillation solution to the
atmospheric neutrino problem. The analysis refers to a global fit
to all the available data. In the case of the sterile channels,
($-$) and ($+$) stand for negative and positive $\Delta m^2$, respectively.}
\vspace{-15pt}
\end{table}
%%%%%%%%%%%%%%%%%%%%%%%%%%%%%%%%%%%%%%%%%%%%%%%%%%%%%%%%%%%%%%%%%%%%%%%%%%%%%%%%%%%%%%%%%%%%%%%

%\vspace{0.5cm}
{\bf Acknowledgements.}
I wish to address my warmest thanks to the TAUP99 Organizers for
allowing me to deliver this second talk to the Conference. 
This work was supported by the Spanish DGICYT under grant number 
PB95--1077 and by the TMR network grant ERBFMRXCT960090 of the European
Union.

\end{document}